\documentclass[conference]{IEEEtran}
\IEEEoverridecommandlockouts
\usepackage{cite}
\usepackage{amsmath,amssymb,amsfonts}
\usepackage{algorithmic}
\usepackage{multirow}
\usepackage{graphicx}
\usepackage{textcomp}
\usepackage{xcolor}
\def\BibTeX{{\rm B\kern-.05em{\sc i\kern-.025em b}\kern-.08em
    T\kern-.1667em\lower.7ex\hbox{E}\kern-.125emX}}
\begin{document}

\title{A Polymorphic Electro-Optic Logic Gate for High-Speed Reconfigurable Computing Circuits\\
{}
\thanks{This research is supported by a grant from NSF (CNS-2139167)}
}

\author{\IEEEauthorblockN{Venkata Sai Praneeth Karempudi}
\IEEEauthorblockA{\textit{Department of ECE} \\
\textit{University of Kentucky}\\
Lexington, Kentucky, USA \\
kvspraneeth@uky.edu}
\and
\IEEEauthorblockN{Sairam Sri Vatsavai}
\IEEEauthorblockA{\textit{Department of ECE} \\
\textit{University of Kentucky}\\
Lexington, Kentucky, USA \\
Sairam\_Srivatsavai@uky.edu}
\and
\IEEEauthorblockN{Ishan Thakkar}
\IEEEauthorblockA{\textit{Department of ECE} \\
\textit{University of Kentucky}\\
Lexington, Kentucky, USA \\
igthakkar@uky.edu}
\and
\IEEEauthorblockN{Jeffrey Todd Hastings}
\IEEEauthorblockA{\textit{Department of ECE} \\
\textit{University of Kentucky}\\
Lexington, Kentucky, USA \\
todd.hastings@uky.edu}
}

\maketitle

\begin{abstract}
In the wake of dwindling Moore's law, integrated electro-optic (E-O) computing circuits have shown revolutionary potential to provide progressively faster and more efficient hardware for computing. The E-O circuits for computing from the literature can operate with minimal latency at high bit-rates. However, they face shortcomings due to their operand handling complexity, non-amortizable high area and static power overheads, and general unsuitability for large-scale integration on reticle-limited chips. To alleviate these shortcomings, in this paper, we present a microring resonator (MRR) based polymorphic E-O logic gate (MRR-PEOLG) that can be dynamically programmed to implement different logic functions at different times. Our MRR-PEOLG can provide compactness and polymorphism to E-O circuits, to consequently improve their operand handling and amortization of area and static power overheads. We model our MRR-PEOLG using photonics foundry-validated tools to perform frequency and time-domain analysis of its polymorphic logic functions. Our evaluation shows that the use of our MRR-PEOLG in two E-O circuits from prior works can reduce their area-energy-delay product by up to 82.6$\times$. A tutorial on the modeling and simulation of our MRR-PEOLG, along with related codes and files, is available on https://github.com/uky-UCAT/MRR-PEOLG.
\end{abstract}

\begin{IEEEkeywords}
Polymorphic, Microring Resonator, Temperature, Bit-rate.
\end{IEEEkeywords}

\section{Introduction}
Moore’s law has been steering the advancement of computing hardware since its inception. But unfortunately, in recent years, it has faced fatal challenges as the nanofabrication technology is experiencing physical limitations due to the exceedingly small size of transistors \cite{amlan2022}. This has forced researchers in industry and academia to develop new more-than-Moore technologies that can continue to provide persistently faster and more efficient computing hardware \cite{amlan2022}. Fortunately, silicon photonics (SiP) enabled electro-optic (E-O) circuit integration has been identified as one such promising technology \cite{ying2020}. The SiP-based E-O circuits are generally CMOS compatible and provide several advantages over their purely electrical counterparts. These advantages include sub-picosecond speeds, low dynamic power consumption and distance-independent bit-rate \cite{ying2020}. Due to these advantages compared to the CMOS-based electrical circuits, the early prototypes of SiP-based E-O circuits for computing (e.g., \cite{lightbulb,shiflett2020,shiflett2021,qiu2012,ying2019integrated,karempudi2021,ying2020}) have been shown to provide up to two orders of magnitude improvements in performance and energy efficiency \cite{bangari2019}\cite{ying2020}. 

The SiP-based E-O circuits for computing, which have been demonstrated in prior works (e.g., \cite{lightbulb,shiflett2020,shiflett2021,qiu2012,ying2019integrated,karempudi2021,ying2020}) are typically used to implement the following four types of logical and arithmetic functions: \textbf{(I)} \textbf{Basic logic-gate functions.} For instance, a microring resonator (MRR) integrated phase change memory (PCM) device based XNOR gate is employed in \cite{lightbulb} to enable acceleration of binary neural networks. Similarly, in \cite{shiflett2020} and \cite{shiflett2021}, an add-drop MRR based AND gate is employed to enable partial multiplications of two binary operands, to aid the acceleration of deep neural networks. \textbf{(II) Arbitrary combinational logic functions.} For example, the directed logic based MRR-enabled reconfigurable E-O circuits are demonstrated in \cite{qiu2012} and \cite{ying2019integrated}. These can work as the direct optical replacement of field programmable gate arrays (FPGAs). \textbf{(III) Two-operand arithmetic functions.} High-speed E-O circuits for partial sum accumulation and two-operand addition have been demonstrated (e.g., \cite{karempudi2021}, \cite{ying2020}) with various designs supporting custom precision \cite{karempudi2021} and full-precision polymorphic operation \cite{ying2020}. \textbf{(IV) Multi-operand linear arithmetic functions.} Several analog and digital E-O circuits based on MRRs and/or Mach-Zehnder Interferometers (MZIs) have been demonstrated to implement Multiply-Accumulate (MAC) and Vector Dot Product (VDP) operations (e.g., \cite{lightbulb,shiflett2020,liu2019,bangari2019}) for deep learning workloads. These logical and arithmetic functions implemented using E-O circuits typically fulfill the requirements of ultra-fast, highly-parallel general purpose computing or deep learning acceleration.  

However, we observe that these SiP-based E-O circuits from prior works face three major shortcomings. \textit{First}, the E-O circuits for simple logic-gate functions intake the two input operands differently; one operand is typically applied optically and the other operand is applied electrically. For instance, in the E-O XNOR gate from \cite{lightbulb} and the E-O AND gate from \cite{shiflett2020}, one of the two operands has to be modulated onto the incoming optical wavelengths, for which an additional optical modulator device per gate function is required, assuming that the utilized laser sources provide unmodulated optical power. Having to provide one of the operands optically through an additional modulator device increases the hardware area overhead and the operand handling complexity in the E-O circuits. Instead, there is a need to design a simpler hardware, which can be achieved by promoting all electrical provisioning of both the operands. \textit{Second}, the E-O circuits for arithmetic functions occupy very large areas compared to CMOS implementations. For instance, the E-O MAC circuit used in \cite{shiflett2021} occupies up to 100$\times$ more area compared to the all-electric MAC circuit \cite{shiflett2021}. Moreover, such E-O circuits for arithmetic functions can hardly achieve more than 60\% hardware utilization \cite{shiflett2020}. This is because these circuits typically belong to larger processing units where they occupy only part of the entire end-to-end datapath \cite{bangari2019,lightbulb,liu2019}. Such low hardware utilization often leads to high idle time and consequently very high, non-amortizable area and static power overheads. This in turn motivates the need for more flexible E-O circuits that can adapt to different arithmetic/logic functions at different times, to increase the amortization of their high area and static power overheads by reducing their total idle time. \textit{Third}, the high area overhead of E-O circuits makes them less suitable for highly-parallel Single-Instruction-Multiple-Data (SIMD), Multiple-Instruction-Multiple-Data (MIMD), and Systolic Array (SA) based processing architectures. This is because SIMD, MIMD and SA architectures typically employ thousands of streaming processing units, with each processing unit requiring multiple copies of basic logical and arithmetic functions. Implementing these functions using bulky E-O circuits with 100$\times$ more area can drastically reduce the number of processing units that can be integrated on a single chip whose area is typically limited by the reticle size ($<=$900 mm$^2$ \cite{naffziger2021}). Since SIMD, MIMD, and SA based processing units have become extremely popular for executing modern Euclidean as well as non-Euclidean data workloads (i.e., workloads with grid and graph structured data), it becomes crucial to alleviate the unsuitability of E-O circuits for SIMD, MIMD and SA based designs by forging new E-O circuits with relatively low area overheads. 

To address these shortcomings, in this paper, we present a single \underline{\textbf{MRR}} based \underline{\textbf{P}}olymorphic \underline{\textbf{E-O}} \underline{\textbf{L}}ogic \underline{\textbf{G}}ate (\textbf{MRR-PEOLG}). Our MRR-PEOLG can accept both input operands electrically, and its drop-port (through-port) optical response can be thermo-optically programmed to make it dynamically follow the truth table of different logic functions, such as AND, OR and XOR (NAND, NOR, and XNOR), at different times. Consequently, the E-O circuits built using our MRR-PEOLG can address the above-described shortcomings by providing (1) the ability of all-electrical application of the input operands, (2) compactness through a single-MRR structure of the E-O gate, and (3) high flexibility through the introduced polymorphism, and consequently, low idle time and improved suitability for use with SIMD/MIMD/SA based architectures.

The key contributions of this paper are summarized below: 

\begin{itemize}
\item We model our MRR-PEOLG using the photonics foundry-validated tools from Ansys/Lumerical \cite{lumericalwebsite}, and then, perform the frequency, time-domain transient, and thermal analysis for different logic-gate functions. A tutorial on the modeling and simulation of our MRR-PEOLG, along with related codes and files, is available on https://github.com/uky-UCAT/MRR-PEOLG;
\item Based on our analysis, we evaluate the performance of our MRR-PEOLG, from which we determine the maximum achievable bit-rate and thermal tuning power for each logic-gate function supported by our MRR-PEOLG;
\item We show that the use of our MRR-PEOLG in two E-O circuits from prior works can provide improvement in area-energy-delay product of up to 82.6$\times$;
\item We also discuss how MRR-PEOLG can be used to realize E-O reconfigurable SIMD/MIMD architectures. 
\end{itemize}

\begin{figure}[h!]
    \centering
    \includegraphics[width = 20pc]{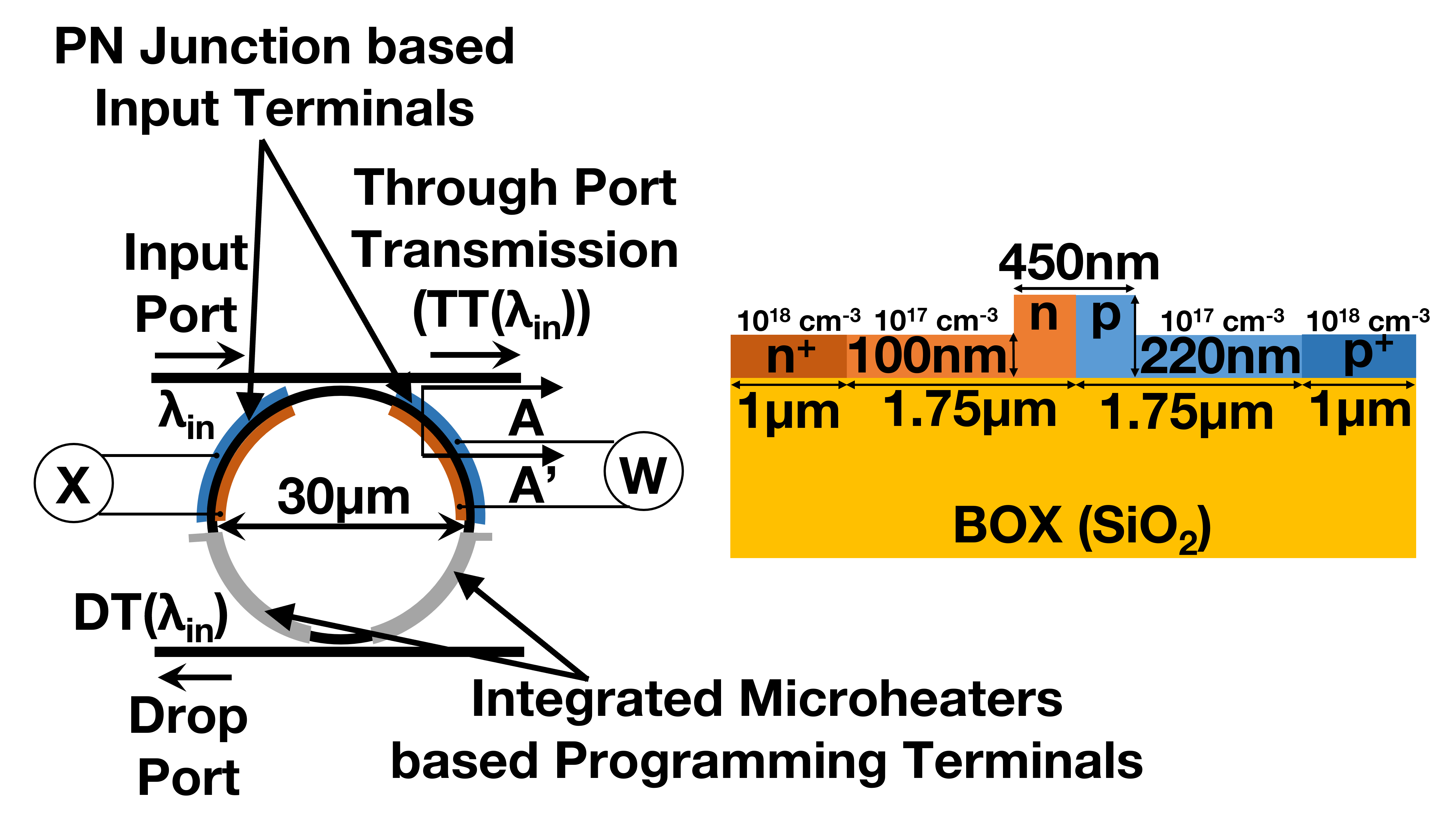}
    \caption{Structure and cross-section of our MRR based polymorphic E-O logic gate (MRR-PEOLG).}
    \label{Fig:1}
\end{figure}

\section{MRR-Based Polymorphic Electro-Optic Logic Gate (MRR-PEOLG)}
\subsection{Structure}
Our MRR-PEOLG is basically an add-drop MRR \cite{bogaerts2012} with four quarter-sized phase-shifting sections embedded in it, as shown in Fig. \ref{Fig:1}(a). Two quarter-sized sections of the MRR are two PN junctions which are operated in the forward bias condition, whereas the remaining two quarter-sized sections integrate micro-heaters. The cross-section of a PN-junction based section of our MRR-PEOLG is shown in the right hand side of Fig. \ref{Fig:1}(a), which consists of a ridge waveguide with an embedded lateral PN junction, fabricated on the top of a buried oxide layer. The dimensions of the P-type and N-type regions, and their corresponding carrier concentrations are also provided in Fig. \ref{Fig:1}(a). The PN junction based sections of our MRR-PEOLG work as the input terminals where the input logic signals/operand bits are applied. On the other hand, the microheaters integrated sections of our MRR-PEOLG work as the programming terminals that are used to program the MRR-PEOLG to perform specific logic-gate functions.

Applying a voltage to the microheaters based programming terminals can increase the temperature of the MRR, which in turn can shift (red shift) the resonance of the MRR towards the longer wavelength. This is because of the thermo-optic effect in silicon (\cite{bahadori2017}). To program MRR-PEOLG to implement a specific logic-gate function, the operand-independent MRR resonance (i.e., the programmed MRR resonance) is adjusted to a specific spectral position with respect to the input optical wavelength, by applying a voltage to the programming terminals. Then, the electrical input logic signals or input operand bits (x and w) are applied to the PN junctions based input terminals of the MRR. Upon doing so, the resonance of the MRR shifts (blue shifts) towards the shorter wavelength depending on the combination of the applied input operand bits. This is because of the free-carrier plasma dispersion effect in silicon (\cite{mulcahy2022}). Applying the input operand bits to the input terminals makes the through-port and drop-port optical responses of our MRR-PEOLG follow the truth-table of the logic-gate functions for which the MRR-PEOLG is programmed. In this manner, our MRR-PEOLG can perform different logic-gate functions at different times. At any given time, the through-port optical response of the MRR-PEOLG follows logical complement of the drop-port optical response. Therefore, AND, OR and XOR functions can be realized (one function at a time) at the drop port of the MRR-PEOLG. Concurrently, the through port of the MRR-PEOLG can provide complementary logic-gate functions such as NAND, NOR and XNOR as discussed below.

\subsection{Modeling}
\begin{figure}[h!]
    \centering
    \includegraphics[width = 20pc]{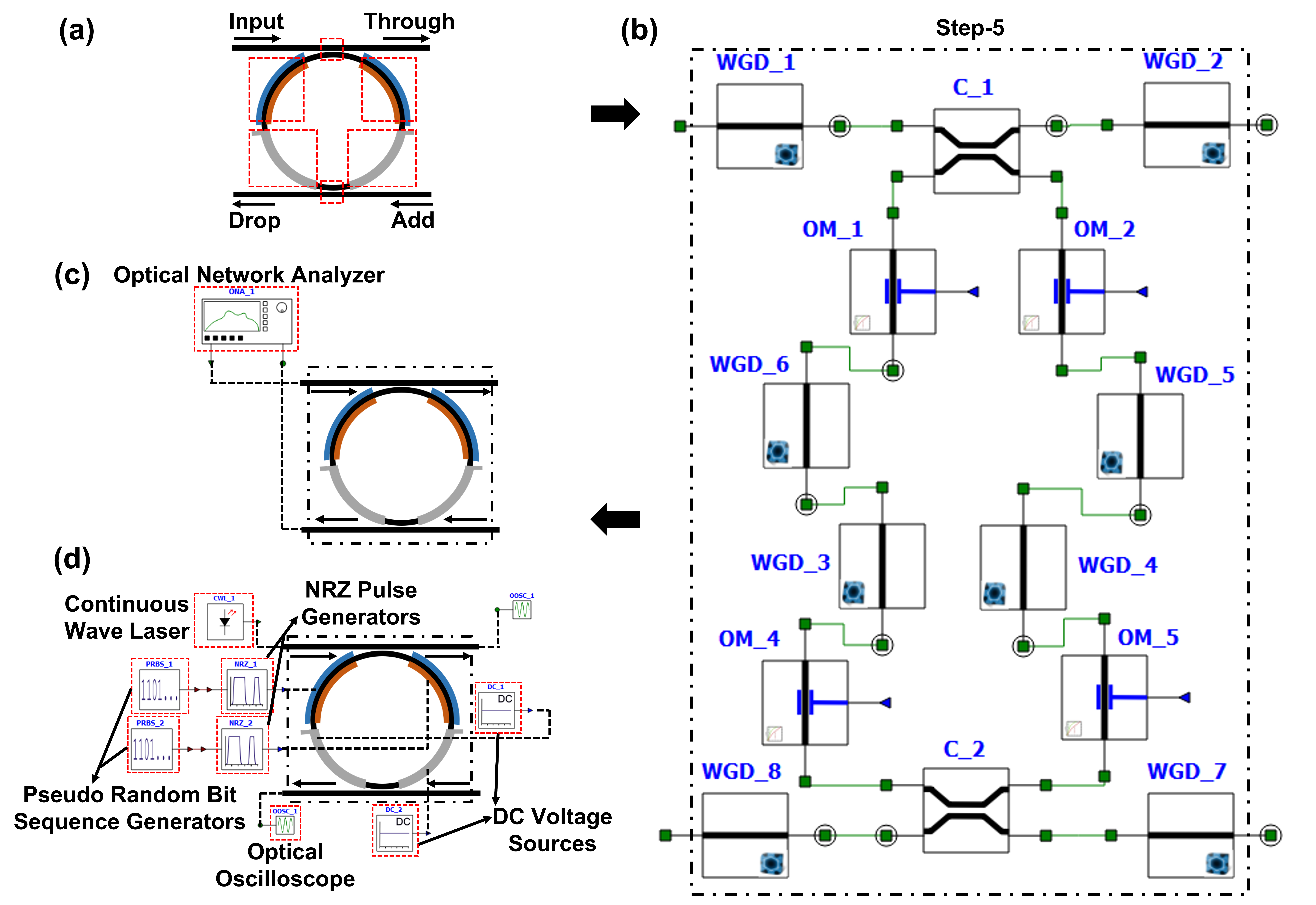}
    \caption{(a) Step-1 to Step-4, and (b) Step-5 of the procedure used for modeling our MRR-PEOLG. The schematic simulation setup in  ANSYS/Lumerical's INTERCONNECT tool for (c) frequency-domain and (d) time-domain transient analysis of our MRR-PEOLG.}
    \label{Fig:2}
\end{figure}

We model our MRR-PEOLG using the photonics foundry-validated simulation tools from Ansys/Lumerical \cite{lumericalwebsite}. We break down our MRR-PEOLG design into a set of primitive elements. Fig. \ref{Fig:2}(a) shows a schematic of our MRR-PEOLG, whose breakdown into the primitive elements is shown in Fig. \ref{Fig:2}(b). We use different solvers in the Ansys/Lumerical tools \cite{lumericalwebsite} to model each primitive element. From these models, we extract various parameters for each primitive element. Later, we combine all of the extracted parameters in Ansys/Lumerical's INTERCONNECT tool \cite{lumericalwebsite} (tool for the modeling and simulations of photonic integrated circuits) to create our MRR-PEOLG in Fig. \ref{Fig:2}(b).  Finally, we perform the frequency-domain and time-domain transient simulations of our MRR-PEOLG. Different steps for the modeling and simulation of our MRR-PEOLG using the ANSYS/Lumerical tools/solvers are summarized below.

\textbf{Step-1 - modeling MRR-waveguide coupling sections:} First, create coupling sections in the finite difference time domain (FDTD) solver and  extract the power coupling coefficients as a function of wavelength for the fundamental TE mode.  Import these coefficients in the coupling elements C\_1 and C\_2 (Fig. \ref{Fig:2}(b)).   

\textbf{Step-2 - modeling straight waveguide sections:} First, characterize the passive, straight, channel waveguides of the MRR-PEOLG using the finite difference eigenmode (FDE) solver.  Extract the effective index, group index, and dispersion for the waveguides as functions of wavelength.  Load this information into the primitive elements WGD\_1, WGD\_2, WGD\_7, and WGD\_8 (Fig. \ref{Fig:2}(b)).

\textbf{Step-3 - modeling PN-junction based input terminals:} First, create a quarter ring with an embedded lateral PN-junction in the CHARGE tool.  Perform the simulation to extract the spatial distribution of the charge carriers as a function of the bias voltage.
Then, export this data into the FDE solver and calculate the perturbations in the refractive index of the waveguides connected to the input terminals.  Then, calculate the change in the effective index and resonance of the entire MRR-PEOLG as a function of the bias voltage. Import this information into the primitive elements WGD\_6 (connected to OM\_1) and WGD\_5 (connected to OM\_2) (Fig. \ref{Fig:2}(b)).

\textbf{Step-4 - modeling microheaters based programming terminals:} Extract the temperature profile of the MRR-PEOLG as a function of the applied microheater voltage. Then, import this data into the FDE solver to calculate the change in the effective index of the MRR-PEOLG as a function of its temperature. Import this information into the primitive elements WGD\_3 (connected to OM\_4) and WGD\_4 (connected to OM\_5) (Fig. \ref{Fig:2}(b)).

\textbf{Step-5 - preparing for simulations:} Connect the primitive-elements based model of the MRR-PEOLG (Fig. \ref{Fig:2}(b)) with other testing and characterization apparatus in the INTERCONNECT tool, as shown in Fig. \ref{Fig:2}(c) and Fig. \ref{Fig:2}(d).

\subsection{Operation}
\begin{figure*}[h!]
    \centering
    \includegraphics[width = 42pc]{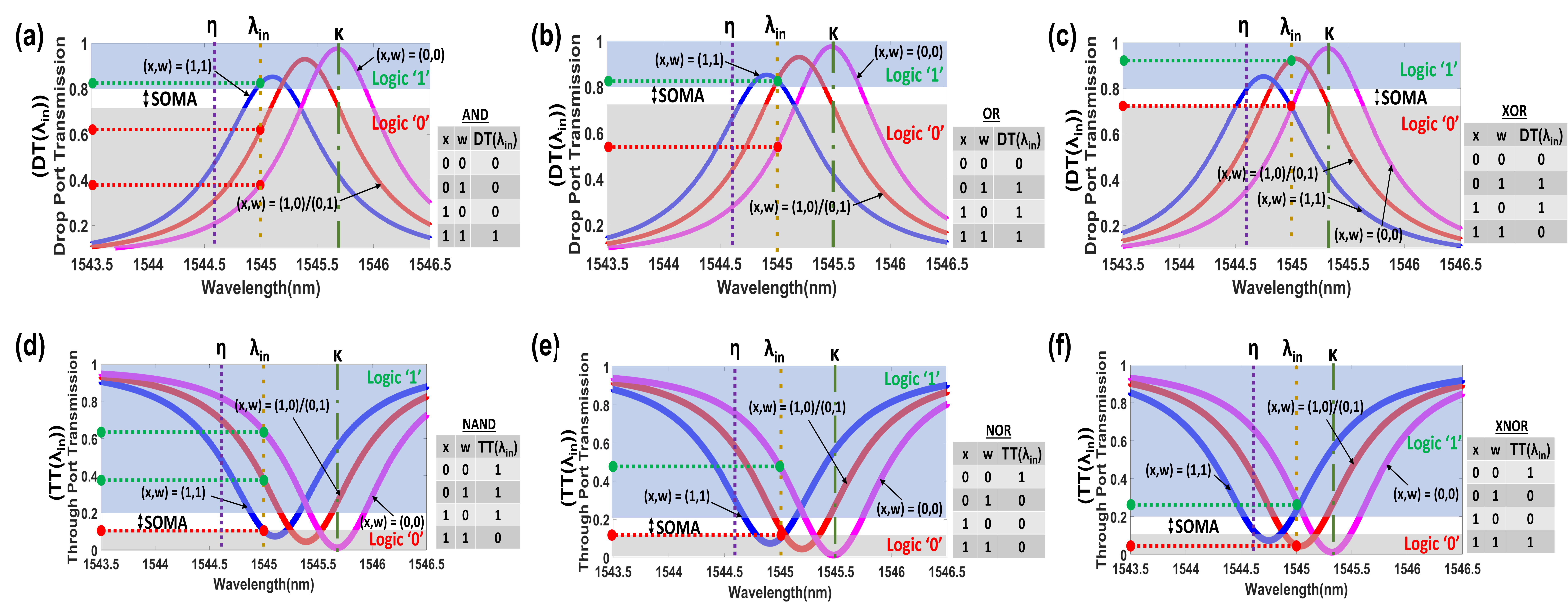}
    \caption{The transmission spectra obtained at the drop port of our MRR-PEOLG for logic-gate functions (a) AND, (b) OR, and (c) XOR, and at the through port of our MRR-PEOLG for complementary logic-gate functions (d) NAND, (e) NOR, and (f) XNOR.}
    \label{Fig:3}
\end{figure*}

To explain the operation of our MRR-PEOLG, we performed frequency-domain simulations using the INTERCONNECT tool \cite{lumericalwebsite}. Our simulation setup for this frequency-domain analysis is shown in Fig. \ref{Fig:2}(c). Accordingly, we connected an optical network analyzer (ONA) to our MRR-PEOLG to extract the transmission spectra at its drop and through ports. We extracted the transmission spectra for different values of the detuning of the operand-independent MRR resonance position $\kappa$ with respect to the input wavelength $\lambda_{in}$. As mentioned earlier, these detuning values correspond to different logic-gate functions that the MRR-PEOLG can perform. In addition, we also extracted transmission spectra for different combinations of the input operand bits. All of these transmission spectra for different logic-gate and complementary logic-gate functions are shown in Figs. \ref{Fig:3}(a) to \ref{Fig:3}(f). Transmission spectra corresponding to logic-gate functions AND, OR and XOR are shown in Figs. \ref{Fig:3}(a), \ref{Fig:3}(b), and \ref{Fig:3}(c) respectively. These transmission spectra are drop-port transmission spectra (Lorentzian lineshape passbands). Similarly, the transmission spectra corresponding to complementary logic-gate functions NAND, NOR and XNOR are shown in Figs. \ref{Fig:3}(d), \ref{Fig:3}(e), and \ref{Fig:3}(f) respectively. These transmission spectra are through-port transmission spectra (inverse Lorentzian lineshape passbands). As we can see in Fig. \ref{Fig:3}, the drop port and through port transmission exhibits two clearly distinguishable levels. The full transmission range at the drop port and through port of our MRR-PEOLG is divided into two areas, in which the lower part of the full transmission range is indicated with shaded gray whereas the upper part is indicated with shaded blue. If the drop port (DT($\lambda_{in}$)) and through port (TT($\lambda_{in}$)) transmission at $\lambda_{in}$ falls in the lower part of the full transmission range (i.e., in the gray-shaded area), then it is referred to as logic ‘0’ transmission. On the other hand, if the drop port and through port transmission at $\lambda_{in}$ falls in the upper part of the full transmission range (i.e., in the blue-shaded area), then it is referred to as logic ‘1’ transmission. However, the vertical spans of the two distinguishable transmission levels differ between the drop port and through port. This is because, similar to the transmission spectra (Fig. \ref{Fig:3}), the spans of transmission levels at the drop port also complement the spans of transmission levels at the through port. The difference between the minimum supported logic ‘1’ transmission and the maximum supported logic ‘0’ transmission is the sensitivity of optical modulation amplitude (SOMA). SOMA is a property of the photodetector based receiver circuit, and it affects the performance of the MRR-PEOLG (as will be discussed in Sections III and IV). 

To clearly understand the operation of our MRR-PEOLG, let us consider the example of AND function, as shown in Fig. \ref{Fig:3}(a). To program our MRR-PEOLG to implement AND function, a 0.9 V voltage (3.52 mW power) is applied to the programming terminals of the MRR-PEOLG. This shifts the resonance from the initial position, $\eta$, to the programmed position, $\kappa$, where $\kappa$ has the programmed detuning of 0.7 nm with respect to $\lambda_{in}$. Then, the input operand bits x and w are applied to the input terminals of the device. Doing so induces a blueshift in the MRR resonance, the magnitude of which depends on the specific combination of the applied input operand bits (x and w, as shown in Fig. \ref{Fig:3}(a)). If the applied bit-combination (x,w) is (0,0), the resonance position of the MRR stays at $\kappa$ (magenta colored passband in Fig. \ref{Fig:3}(a)) and the drop port transmission at $\lambda_{in}$ provides logic '0' level (the bottom red dot on the Y-axis). If the applied bit-combination (x,w) is (0,1) or (1,0), the position of the MRR resonance changes (red/orange colored passband in Fig. \ref{Fig:3}(a)), but the blueshift is the same for both (0,1) and (1,0) bit combinations, and the drop port transmission at $\lambda_{in}$ still remains at logic '0' level (the top red dot on the Y-axis). On the other hand, if the applied bit-combination (x,w) is (1,1), the MRR resonance undergoes a larger blueshift (blue colored passband in Fig. \ref{Fig:3}(a)), and the position of the passband with respect to $\lambda_{in}$ changes. As a result, the drop port transmission at $\lambda_{in}$ changes to logic '1' level (the green dot on the Y-axis). Hence, the drop port transmission at $\lambda_{in}$ for our MRR-PEOLG changes with the applied input operand bits, and follows the truth table of the AND logic function (see the truth table in Fig. \ref{Fig:3}(a)). As discussed earlier, since the through-port response provides a logical complement to the drop-port response, this AND function at the drop port of the MRR-PEOLG corresponds to NAND function at the through port of the MRR-PEOLG as illustrated in Fig. \ref{Fig:3}(d). 

Similarly, our MRR-PEOLG can be reconfigured to implement OR (NOR) and XOR (XNOR) gate functions as well, by applying a suitable voltage to the programming terminals of our MRR-PEOLG to set the relative position of $\kappa$ with respect to $\lambda_{in}$ as shown in Figs. \ref{Fig:3}(b) and \ref{Fig:3}(c) (transmission spectra corresponding to NOR and XNOR are shown in figs. \ref{Fig:3}(e) and \ref{Fig:3}(f) respectively). Table \ref{Table:1} provides the total power consumed in the microheaters, the programmed detuning ($\kappa$-$\lambda_{in}$), and the required resonance shifting ($\eta$-$\kappa$), to program our MRR-PEOLG for implementing various logic functions. From Table \ref{Table:1}, the power consumed in the  microheaters is proportional to the required resonance shifting ($\eta$-$\kappa$).

\begin{table}
\centering
\caption{Power consumed in the microheaters, programmed detuning, and required resonance shifting, used to program our MRR-PEOLG for implementing different logic functions.}
\begin{tabular}{|c|c|c|c|}
\hline
\textbf{\begin{tabular}[c]{@{}c@{}}Logic-Gate \\ Functions \end{tabular}} &  \textbf{\begin{tabular}[c]{@{}c@{}}Microheater \\ Power (mW)\end{tabular}} & \textbf{\begin{tabular}[c]{@{}c@{}}Programmed\\ Detuning\\ ($\kappa$-$\lambda_{in}$) (nm)\end{tabular}} & \textbf{\begin{tabular}[c]{@{}c@{}}Required\\ Shifting\\ ($\eta$-$\kappa$) (nm)\end{tabular}} \\ \hline
AND / NAND                                                                 & 3.52                                                                           & 0.7                                                                                                        & -1.1                                                                                             \\ \hline
OR / NOR                                                                                                                             & 2.93                                                                           & 0.5                                                                                                       & -0.9                                                                                             \\ \hline

XOR / XNOR                                                                                                                                  & 2.3                                                                            & 0.3                                                                                                       & -0.7                                                                                             \\ \hline
\end{tabular}
\label{Table:1}
\end{table}

\section{Transient Analysis}
\subsection{Method}
\begin{figure}[h!]
    \centering
    \includegraphics[width = 21pc]{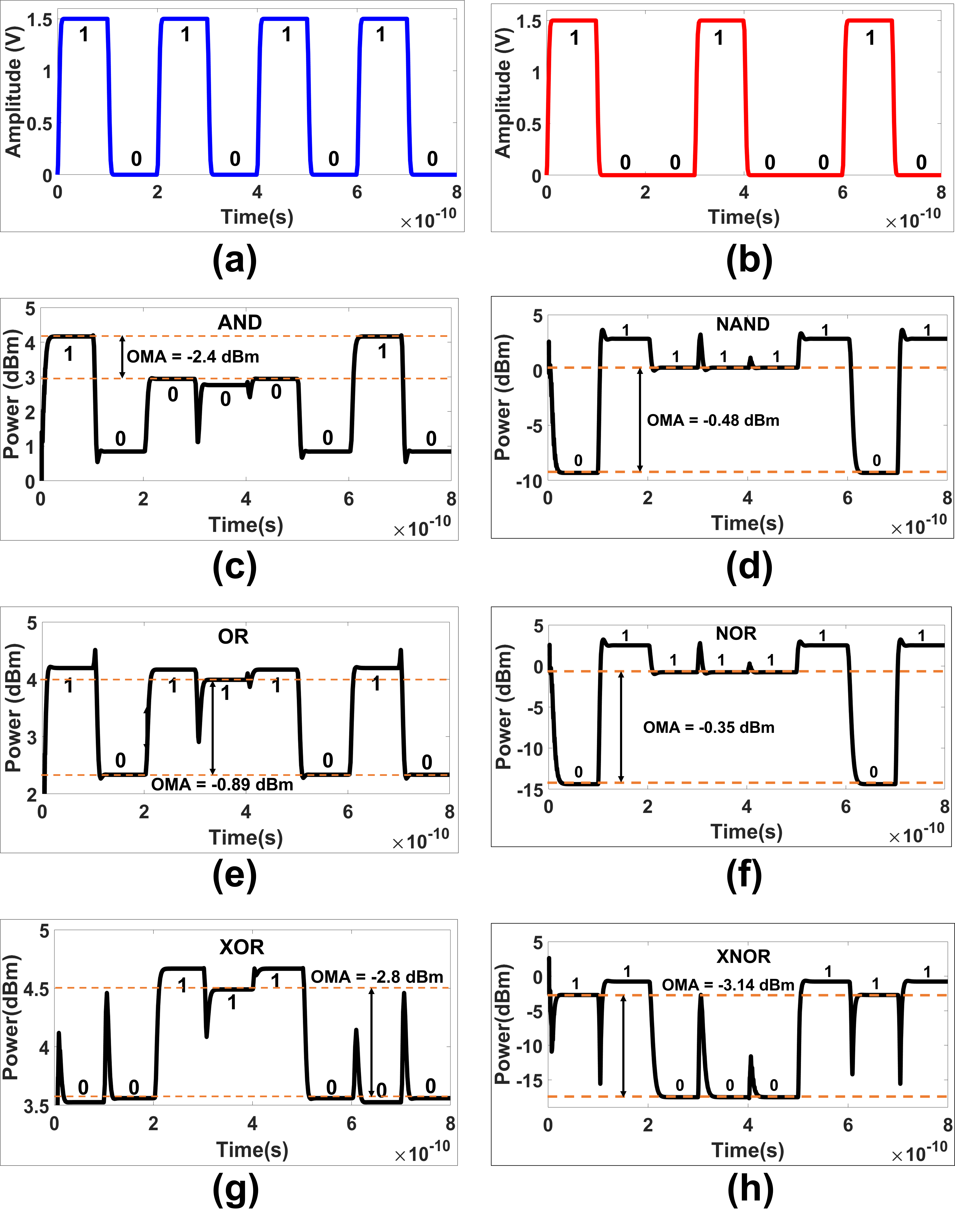}
    \caption{(a),(b) The electrical pulse signals of 10 Gb/s bit-rate provided as input to the PN junctions of our MRR-PEOLG. The corresponding output pulse patterns obtained at the drop port of our MRR-PEOLG for logic-gate functions (c) AND, (e) OR, and (g) XOR , and at the through port of our MRR-PEOLG for complementary logic-gate functions (d) NAND, (f) NOR, and (h) XNOR. The optical input power is 5 dBm in all cases.}
    \label{Fig:4}
\end{figure}

As illustrated in Fig. \ref{Fig:2}(d), to perform transient analysis of our MRR-PEOLG in the INTERCONNECT tool, we connected a pseudo random bit sequence (PRBS) generator and a non-return-to-zero (NRZ) pulse generator to each of the input terminals of the MRR-PEOLG. Each PRBS generator generates a random bit sequence of 10 Gb/s, which is given as input to the NRZ pulse generator. Each NRZ pulse generator then generates a sequence of electrical NRZ pulses of 1.5 V amplitude at 10 Gb/s. The blue and red pulses shown in Figs. \ref{Fig:4}(a) and \ref{Fig:4}(b) respectively are the electrical NRZ pulses that we have provided as inputs to the two input terminals of our MRR-PEOLG for the transient analysis. We have also connected a continuous wave (CW) laser to the input port of the MRR-PEOLG, which generates an optical signal of wavelength 1545 nm ($\lambda_{in}$ = 1545 nm in Fig. \ref{Fig:3}) with an optical power of 5 dBm.  We connected optical oscilloscopes to the drop and through ports to record the output pulse patterns corresponding to different logic functions for the given input electrical pulse signals. The results obtained from this transient analysis are discussed next.

\subsection{Results and Discussion}
Fig. \ref{Fig:4}(c), Fig. \ref{Fig:4}(e), and Fig. \ref{Fig:4}(g) illustrate the output pulse signals obtained at the drop-port of the MRR-PEOLG for different logic functions. Similarly, Fig. \ref{Fig:4}(d), Fig. \ref{Fig:4}(f), and Fig. \ref{Fig:4}(h) illustrate the output pulse signals simultaneously obtained at the through-port of the MRR-PEOLG for different complementary logic functions. To obtain these pulse patterns, we first reconfigured the MRR-PEOLG to implement various logic functions by changing the temperature using the integrated microheaters. We then followed the method described in Section III.A. As evident from Fig. \ref{Fig:4}, the output pulse signals follow the pulse-wise truth-tables of the respective logic functions, which confirms the capability of our MRR-PEOLG to correctly realize different logic functions.

From Figs. \ref{Fig:4}(c) - \ref{Fig:4}(h), the optical modulation amplitude (OMA), which is the difference between the minimum logic '1' power level and the maximum logic '0' power level in an output pulse pattern, differs for different logic functions. To clearly understand this, let us consider AND, XOR, and OR functions. For the AND function shown in Fig. \ref{Fig:4}(c), the OMA is $\sim$-2.4dBm. This is because, as can be observed from Fig. \ref{Fig:4}(c), the drop port transmission at $\lambda_{in}$ corresponding to the logic ‘1’ output level (i.e., (x,w) = (1,1)) is $\sim$0.82 (the green dot on the Y-axis), whereas the maximum drop port transmission at $\lambda_{in}$, corresponding to the logic ‘0’ output level (i.e., (x,w) = (1,0) or (0,1)), is $\sim$0.62 (the top red dot on the Y-axis). Hence, the OMA, i.e., the difference between the logic ‘1’ optical power level (0.82$\times$5dBm = 2.57 mW) and the logic ‘0’ optical power level(0.62$\times$5dBm = 1.995 mW) is $\sim$-2.4 dBm ($\sim$0.575 mW). Similarly, for OR (NOR) and XOR (XNOR) functions shown in Fig. \ref{Fig:4}, the green and red dots on Y-axis occur at different positions compared to AND (NAND) function. Therefore, our MRR-PEOLG exhibits different OMA for different logic functions.

Since the OMA for the output pulse pattern basically defines how well the logic '1's are distinguishable from logic '0's, having different OMA values renders different reliability bounds for different logic functions implemented by our MRR-PEOLG. In general, to achieve higher reliability without trying to quantify its value, it is desirable to increase the OMA of an output pulse pattern, which can be done in two ways. First, OMA can be increased by increasing the input optical power at $\lambda_{in}$. (We considered an input optical power of 5 dBm for our results discussed in previous paragraph). Second, OMA can be increased by decreasing the full width at half maximum (FWHM) or 3-dB bandwidth of the MRR-PEOLG. A lower FWHM would make the roll-off edges of the MRR passbands corresponding to (0,0), (0,1)/(1,0) and (1,1) steeper (Fig. \ref{Fig:3}), which in turn would increase the distance between the green dot and top red dot on the Y-axis, thereby increasing the OMA. Note that increasing OMA is not always necessary, as a low OMA would cause reliability issues only if it is lower than the OMA sensitivity (SOMA) of the receiver circuit that is employed to make sense of the output pulse pattern. Therefore, decreasing SOMA of the receiver circuit can also increase the reliability of our MRR-PEOLG. Thus, the FWHM (3-dB bandwidth) of the MRR-PEOLG, the SOMA of the receiver circuit, and the input power are the three factors that influence the impact of OMA on the reliability of our MRR-PEOLG.

\begin{figure}[h!]
\centering
    \includegraphics[width = 21pc]{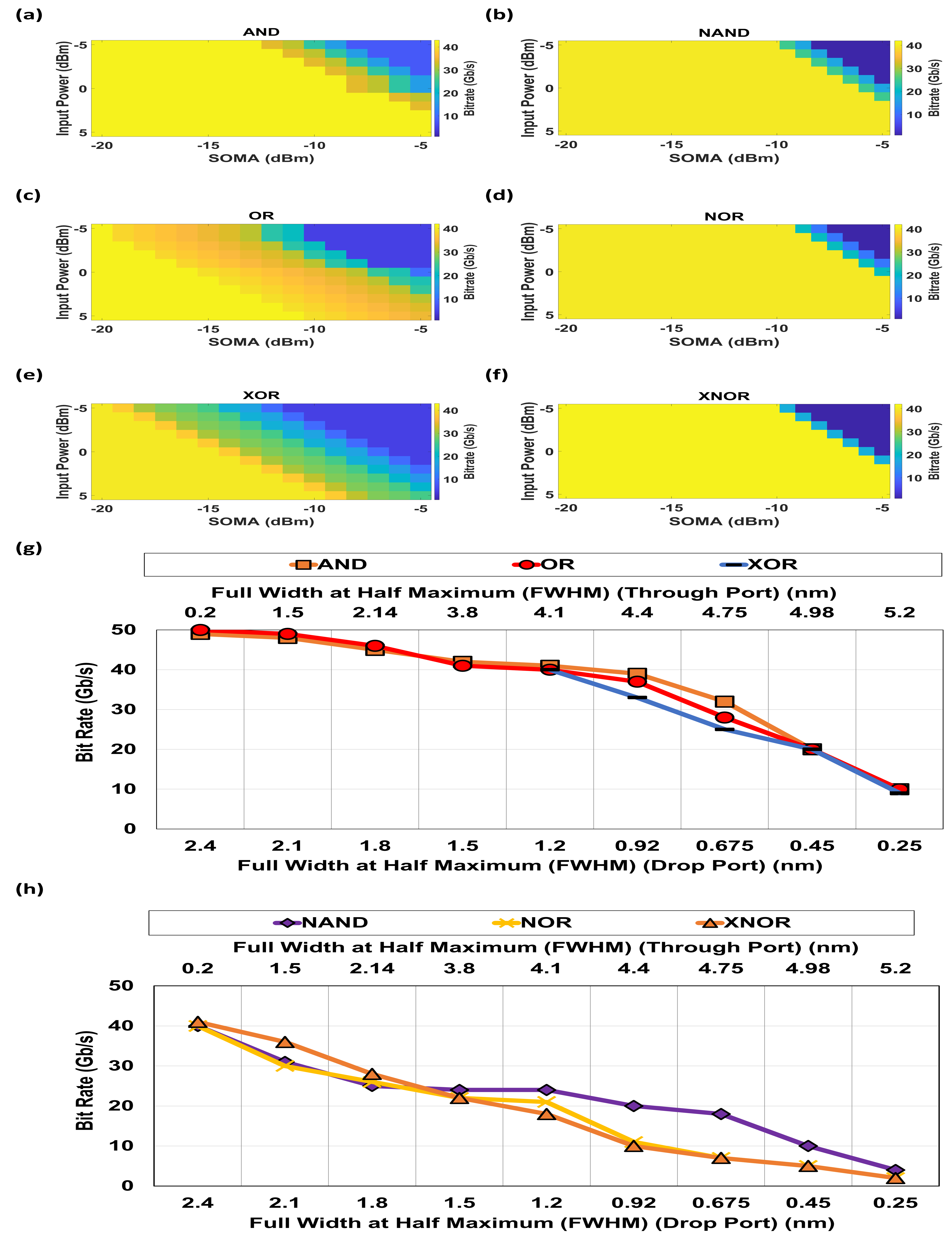}
    \caption{Colormap plots for logic functions (a) AND, (c) OR, (e) XOR  (obtained at the drop port of our MRR-PEOLG), and complementary logic functions (b) NAND, (d) NOR, (f) XNOR  (obtained at the through port of our MRR-PEOLG) that depict the maximum achievable bit-rate for given input optical power and SOMA. These color maps are evaluated for drop-port FWHM of 1.2 nm. We also report the maximum achievable bit-rate corresponding to (g) AND, OR, and XOR functions, and (h) NAND, NOR, and XNOR functions, evaluated for different values of FWHM, 0 dBm input optical power, and -5 dBm SOMA.}
    \label{Fig:5}
\end{figure}

Moreover, these three factors impact the maximum speed (bit-rate) at which the input pulse patterns can be driven. Increasing the bit-rate will reduce OMA because either the free-carrier concentration in the PN junctions or the optical energy inside the MRR does not change as fast as the applied input electrical pulse signals. For a given FWHM (3-dB bandwidth), it is possible to keep increasing the bit-rate until the OMA becomes less than the SOMA limit of the receiver circuit. Once the OMA for a given input power crosses the SOMA limit, the OMA can be increased to support a higher bit-rate by increasing the input power. Therefore, the maximum achievable bit-rate for our MRR-PEOLG depends on SOMA, FWHM, and input optical power. We have evaluated the maximum achievable bit-rate for our MRR-PEOLG, corresponding to various logic functions, which is discussed in next section.

\section{Performance Analysis}
For this analysis, we have used the scripting capabilities available in the ANSYS/Lumerical tools to run a performance evaluation of our MRR-PEOLG. We swept the input optical power in the range from -5 dBm to 5 dBm. Similarly, we swept SOMA in the range from -5 dBm to -20 dBm. Then, for each combination of the input optical power and SOMA, we evaluated the maximum achievable bit-rate for each logic-gate function supported by our MRR-PEOLG. The results of this analysis are shown in Fig. \ref{Fig:5} in the form of colormap plots.

\subsection{Results and Discussion}
The colormap plots in Figs. \ref{Fig:5}(a) to \ref{Fig:5}(f) depict the maximum achievable bit-rate corresponding to each logic-gate function for an FWHM of 1.2 nm and different combinations of input optical power and SOMA. From the colormap plots, the AND function achieves a maximum bit-rate of 42 Gb/s across all SOMA values if the input optical power is $>$2 dBm, as well as across all input power values if the SOMA value is $<$-13 dBm. Similarly, OR and XOR functions achieve a maximum bit-rate of 41 Gb/s and 40 Gb/s respectively across all input optical power values if SOMA is $<$-19 dBm. Meanwhile, the NAND function achieves a maximum bit-rate of 40 Gb/s across all SOMA values if the input optical power is $>$2 dBm, as well as across all input power values if the SOMA value is $<$-11 dBm. Similarly, NOR and XNOR functions achieve a maximum bit-rate of 40 Gb/s and 41 Gb/s respectively across all input optical power values if SOMA is $<$-11 dBm. Moreover, we also show in Figs. \ref{Fig:5}(g) and \ref{Fig:5}(h) that increasing the drop-port FWHM (which can be achieved by increasing the cross-coupling co-efficient of the MRR-PEOLG) can increase the maximum achievable bit-rate for each logic function. These results imply that our MRR-PEOLG can be operated at up to 40 Gb/s for each of its supported logic-gate functions.

\section{Comparison and Envisioned Use Cases}
\subsection{Comparison with E-O Circuits from Prior Work}
We evaluated how the use of our MRR-PEOLG impacts the area, latency, and energy consumption of two E-O circuits from prior works \cite{lightbulb} and \cite{shiflett2021}. We replaced the E-O XNOR gates with our MRR-PEOLG in the E-O XNOR-POP circuits of the binary neural network accelerator LightBulb from \cite{lightbulb}. Similarly, we replaced the AND gates with our MRR-PEOLG in the optical bit-serial multiplier circuits of the digital CNN accelerator from \cite{shiflett2021}. As a result, the performance of these E-O circuits substantially improved as shown in Table \ref{table2}. The energy values are the energy per bit values and include the MRR static power as well as laser power. The area and energy benefits in Table \ref{table2} are due to the compactness and better operand handling of our MRR-PEOLG and also our MRR-PEOLG's ability to realize different logic functions with only a single MRR. The latency benefits are due to the fact that our MRR-PEOLG can operate at up to 40 Gb/s, whereas the original bit-serial multiplier circuit from \cite{shiflett2021} can only operate at up to 10 Gb/s. The E-O XNOR-POPCOUNT units from \cite{lightbulb} can operate at a higher bit-rate of 50 Gb/s, but our MRR-PEOLG based variants provide better area-energy-delay product. These results corroborate the excellent capabilities and efficiency benefits of our MRR-PEOLG.

\begin{table}[h!]
\centering
\caption{Performance comparison of E-O circuits. \\ A=Area, E=Energy, L=Latency}
\begin{tabular}{|c|cc|cc|}
\hline
\multirow{2}{*}{\textbf{Metrics}} & \multicolumn{2}{c|}{\textbf{XNOR-POPCOUNT}}             & \multicolumn{2}{c|}{\textbf{Bit-serial Multiplier}}         \\ \cline{2-5} 
                         & \multicolumn{1}{c|}{\cite{lightbulb}} & \textbf{MRR-PEOLG}           & \multicolumn{1}{c|}{\cite{shiflett2021}} & \textbf{MRR-PEOLG}            \\ \hline
A (mm$^2$)                  & \multicolumn{1}{c|}{0.013}    & 0.011 (1.16×)  & \multicolumn{1}{c|}{0.023}    & 0.011 (2.08×)   \\ \hline
E (nJ)                   & \multicolumn{1}{c|}{0.05}     & 0.032 (1.53×)  & \multicolumn{1}{c|}{0.327}    & 0.033 (9.89×)   \\ \hline
L (ns)                   & \multicolumn{1}{c|}{0.02}     & 0.025 (0.8×)   & \multicolumn{1}{c|}{0.1}      & 0.025 (4×)      \\ \hline
A*E*L                    & \multicolumn{1}{c|}{1.3e-5}   & 0.9e-5 (1.44×) & \multicolumn{1}{c|}{75.2e-5}  & 0.91e-5 (82.6×) \\ \hline
\end{tabular}
\label{table2}
\end{table}

\subsection{Envisioned Use Cases for SIMD/MIMD Architectures}
\begin{figure}[h!]
    \centering
    \includegraphics[width = 21pc]{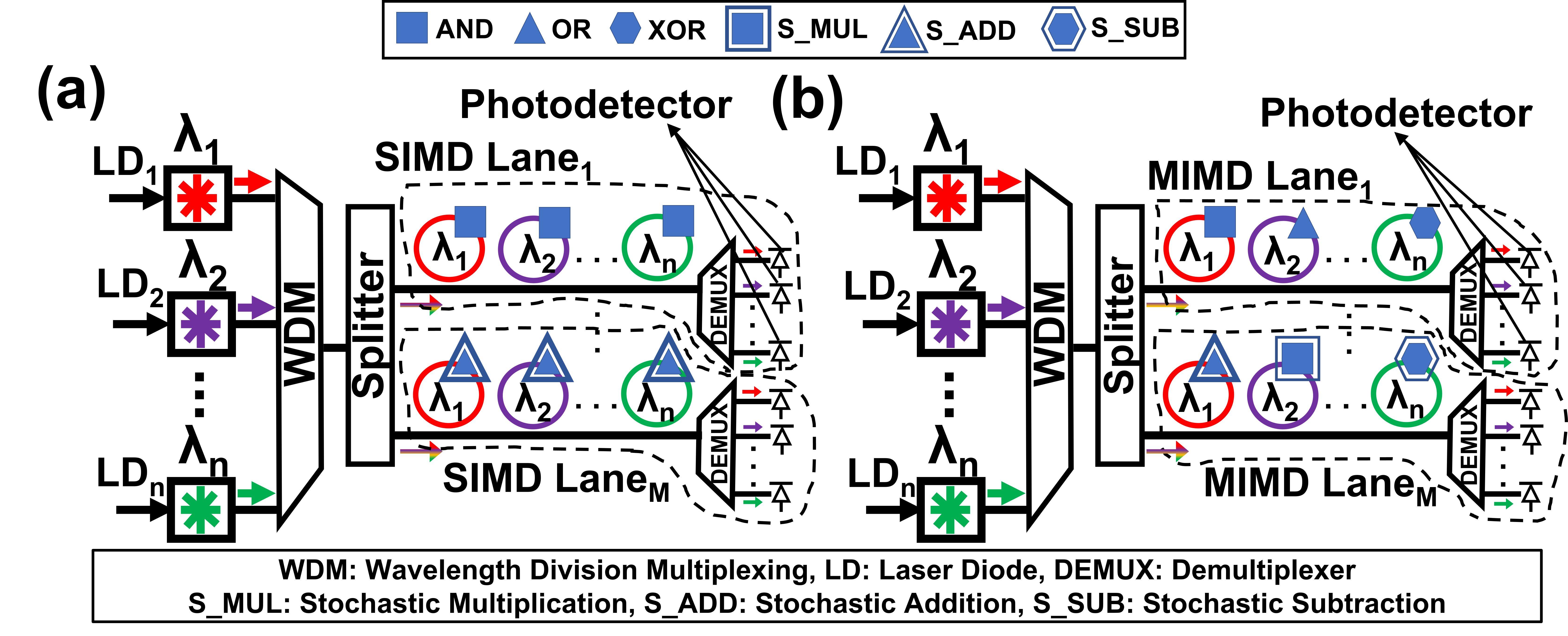}
    \caption{Schematics of how the cascaded arrays of our MRR-PEOLG can be reconfigured to implement (a) a SIMD or (b) an MIMD E-O processing unit. The reconfiguration between SIMD/MIMD can be achieved by programming the individual MRR-PEOLGs for specific logic/arithmetic functions.}
    \label{Fig:6}
\end{figure}

We reason that it is possible to use the dense wavelength division multiplexing (DWDM) technique with our MRR-PEOLG, where cascaded arrays of MRR-PEOLGs can couple with DWDM-enabled rectilinear waveguides. In these cascaded arrays, each MRR-PEOLG can be individually programmed to perform a specific logic-gate function. Moreover, from \cite{wu2020}, it can be inferred that OR, XOR and AND logic-gate functions supported by our MRR-PEOLGs can be useful for realizing stochastic (unary) arithmetic functions such as addition, subtraction and multiplication respectively. This enables the application of the cascaded arrays of MRR-PEOLGs for realizing reconfigurable SIMD/MIMD E-O processing units (see Fig. \ref{Fig:6}). Such E-O SIMD/MIMD units can outperform the traditional GPUs \cite{arunkumar2017} and Tensor Processing Units (TPUs) \cite{wang2019} due to their two-fold benefits. First, they can be operated at higher speeds (up to 40Gb/s) compared to GPUs/TPUs. Second, they can provide significantly better area$\times$latency product, which we plan to evaluate in the future.

\section{Conclusion}
In this paper, we demonstrated a microring resonator based polymorphic electro-optic logic gate (MRR-PEOLG) that can be dynamically reconfigured to implement different logic functions at different times. We modeled our MRR-PEOLG using the photonics foundry-validated simulation tools from ANSYS/Lumerical. Using these tools, we also performed frequency-domain, time-domain transient, and performance analysis of our MRR-PEOLG. From our analysis, we validated that our MRR-PEOLG design can implement various logic functions while operating at speeds of up to 40 Gb/s. Our evaluation shows that the use of our MRR-PEOLG in two E-O circuits from prior works can reduce their area-energy-delay product by up to 82.6$\times$. We also show how our MRR-PEOLG can realize reconfigurable E-O SIMD/MIMD processing units.  

\section*{Acknowledgment}
We thank the anonymous reviewers for their valuable feedback. This research is supported by a grant from NSF (CNS-2139167).

\bibliographystyle{IEEEtran}
\bibliography{ref}

\end{document}